\documentclass[aps,prl,showpacs,twocolumn]{revtex4}  %nofootinbib
\usepackage{amsmath, amsfonts, amsthm, amssymb, graphicx}

\newcommand{\nc}{\newcommand}
\nc{\rnc}{\renewcommand}
\rnc{\d}{\mathrm{d}}
\nc{\D}{\partial}
\rnc{\t}{\tau}
\nc{\K}{\kappa}
\nc{\g}{\gamma}
\nc{\lrarrow}{\leftrightarrow}
\nc{\rg}{\sqrt{g}}
\rnc{\[}{\begin{equation}}
\rnc{\]}{\end{equation}}
\nc{\bea}{\begin{eqnarray}}
\nc{\eea}{\end{eqnarray}}
\nc{\nn}{\nonumber}
\rnc{\(}{\left(}
\rnc{\)}{\right)}
\nc{\dq}{\frac{\d^3 q}{(2\pi)^3}}
\rnc{\a}{\hat{a}}
\nc{\ep}{\epsilon}
\rnc{\tt}{\rightarrow}
\rnc{\inf}{\infty}
\rnc{\Re}{\mathrm{Re}}
\rnc{\Im}{\mathrm{Im}}
\nc{\ie}{{\it i.e.~}}
\nc{\iec}{{\it i.e.,~}}
\nc{\ta}{\bar{a}}
\nc{\vphi}{\varphi}      % use for cosmological scalar field
\nc{\tphi}{\bar{\vphi}}  % use for domain wall scalar field
\nc{\bphi}{\bar{\Phi}}   % use for domain wall perturbation
\nc{\tq}{\bar{q}}
\nc{\tK}{\bar{\K}}
\nc{\tV}{\bar{V}}
\nc{\tr}{\bar{r}}
\rnc{\ta}{\bar{a}}
\nc{\tH}{\bar{H}}
\nc{\tW}{\bar{W}}
\nc{\z}{\zeta}
\nc{\Z}{\mathcal{Z}}
\nc{\W}{\mathcal{W}}
\rnc{\H}{\mathcal{H}}
\nc{\<}{\langle}
\rnc{\>}{\rangle}
\rnc{\O}{\mathcal{O}}
\begin{document}

\title{Holography for Cosmology}

\author{Paul McFadden} 
\email[]{P.L.McFadden@uva.nl}
\author{Kostas Skenderis} 
\email[]{K.Skenderis@uva.nl}

\affiliation{Institute for Theoretical Physics, Valckenierstraat 65, 1018XE Amsterdam.} 

\date{Jan 13, 2010}

\begin{abstract}

  We propose a holographic description of four-dimensional
  single-scalar inflationary universes, and show how cosmological
  observables, such as the primordial power spectrum, are encoded in
  the correlation functions of a three-dimensional quantum field
  theory (QFT).  The holographic description correctly reproduces
  standard inflationary predictions in the regime where a perturbative
  quantization of fluctuations is justified.  In the opposite regime,
  wherein gravity is strongly coupled at early times, we propose a
  holographic description in terms of perturbative large $N$ QFT.
  Initiating a holographic phenomenological approach, we show that
  models containing only two parameters, $N$ and a dimensionful
  coupling constant, are capable of satisfying the current
  observational constraints.

\end{abstract}

%\keywords{}

\maketitle

\paragraph{Introduction.}

Over the last two decades striking new observations have transformed
cosmology from a qualitative to quantitative science. A minimal set of
cosmological parameters characterizing the observed universe, the
concordance cosmology, have now been measured to within a few percent
\cite{Komatsu:2008hk}.  These observations reveal a spatially flat
universe, endowed with small-amplitude primordial perturbations that
are approximately Gaussian and adiabatic with a nearly scale-invariant
spectrum.  These findings are consistent with the generic predictions of
inflationary cosmology and set inflation as the leading theoretical
paradigm for the initial conditions of Big Bang cosmology.  With
future observations promising an unprecedented era
of precision cosmology, the constraints on cosmological parameters are
expected to tighten further still, particularly as regards the
inflationary sector.  This presents a unique window to Planck-scale
physics and a challenge for fundamental theory.

During the last decade we have also witnessed exciting new
developments in fundamental theory.  Holographic dualities have been
proposed and developed leading to a new viewpoint for physical
reality.  Holography states that any quantum theory of gravity should
have a description in terms of a quantum field theory (QFT) which does
not contain gravity in one dimension less.  It is natural to wonder
how cosmology fits into the holographic framework and the main aim of this article 
is to propose a concrete holographic framework for inflationary cosmology.

Apart from the conceptual advances that such a development would
imply, there are also a number of more pragmatic reasons for
developing such a framework. Firstly, uncovering the structure of
three-dimensional QFT in cosmological observables brings in new
intuition about their structure and may lead to more efficient
computational techniques, {\it cf.}~the computation of
non-Gaussianities in \cite{Maldacena:2002vr}.  Secondly, 
holographic dualities are strong/weak coupling dualities meaning that 
in the regime where the one description is weakly coupled the 
other is strongly coupled. This provides an arena for constructing new models
with intrinsic strong-coupling gravitational dynamics at early times
that have only a weakly coupled three-dimensional QFT description
and are thus outside the class of model described by standard inflation.
Such models may well lead to qualitatively different predictions
for the cosmological observables that will be measured in the near future.

Any holographic proposal should specify what the dual QFT is and how
to use it to compute cosmological observables.  The holographic
description we propose uses the one-to-one correspondence between
cosmologies and domain-wall spacetimes discussed in
\cite{Cvetic:1994ya, Skenderis:2006jq} and assumes that the standard
gauge/gravity duality is valid. More precisely, the steps involved are
illustrated in Fig.~1.  The first step is to map any given
inflationary model to a domain-wall spacetime.  For inflationary 
cosmologies that at late times approach either a de Sitter spacetime or a power-law
scaling solution \cite{footnote1},
the corresponding domain-wall solutions describe
holographic renormalization group flows.
For these cases there is an
operational gauge/gravity duality, namely one has a dual description
in terms of a three-dimensional QFT.  Now, the map between cosmologies
and domain-walls can equivalently be expressed entirely in terms of
QFT variables, and amounts to a certain analytic continuation of
parameters and momenta.  Applying this analytic continuation we obtain
the QFT dual of the original cosmological spacetime.

We shall call the resulting theory a `pseudo'-QFT because we currently
only have an operational definition of this theory. Namely, we do the
computations in the QFT theory dual to the corresponding domain-wall
and then apply the analytic continuation.  Perhaps a more fundamental
perspective is to consider the QFT action, with complex parameters and
complex fields as the fundamental objects, and then to consider the
results on different real domains as applicable to either domain-walls
or cosmologies.  Note that the supergravity embedding of the
domain-wall/cosmology correspondence discussed in
\cite{Bergshoeff:2007cg} works in precisely this way.

The holographic description should reproduce standard inflationary
results in their regime of applicability, namely when the fluctuations
around the cosmological background can be perturbatively quantized.
New results should follow by applying the duality in the cases where
such a perturbative quantization of fluctuations is not justified. In
the remainder of this article, we show first that the standard results
are indeed reproduced, then move to construct new models which
are strongly coupled at early times but have a weakly coupled large
$N$ QFT description.

\begin{figure}[tr]
\includegraphics[width=8.5cm]{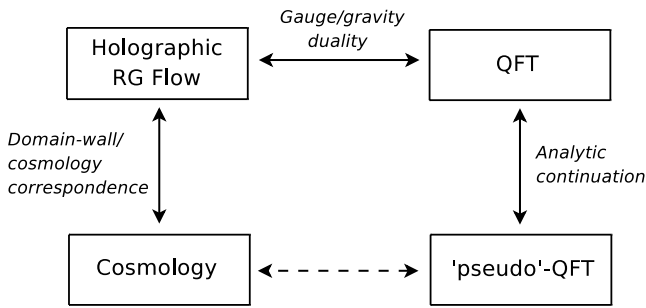} 
\caption{The `pseudo'-QFT dual to inflationary cosmology is operationally defined
using the correspondence of cosmologies to domain-walls and standard gauge/gravity duality.}
\end{figure}

\paragraph{Domain-wall/cosmology correspondence.}

For simplicity, we focus on spatially flat universes equipped 
with a single minimally coupled scalar field, but the results can be extended to
more general cases ({\it eg.}, non-flat, multi-scalar, non-canonical kinetic terms, {\it etc}.)
The linearly perturbed metric and scalar field may be written in the form
\begin{align}
\d s^2 &= \eta \d z^2+a^2(z)[\delta_{ij}+h_{ij}(z,\vec{x})]\d x^i \d x^j, \nn \\
\Phi &= \vphi(z)+\delta\vphi(z,\vec{x}), \label{lin_fl}
\end{align}
where $\eta = -1$ in the case of cosmology, in which case $z$ is the time coordinate,
and $\eta = +1$ in the case of domain-wall solutions, in which case $z$ is the radial
coordinate. We take the domain-wall to be Euclidean for later convenience.  
A Lorentzian domain-wall may be obtained by continuing 
one of the $x^i$ coordinates to become the time coordinate \cite{Skenderis:2006jq}.
The continuation to a Euclidean domain-wall is convenient, however, because the QFT vacuum state implicit in the Euclidean formulation maps to the Bunch-Davies vacuum on the cosmology side.
Other choices of cosmological vacuum require considering the boundary QFT 
in different states, as may be accomplished using the real-time formalism of  \cite{Skenderis:2008dh}. 

The actions for both the cosmology and the Euclidean domain-wall may be written in the combined form
\[
\label{Action}
S = \frac{\eta}{2\K^2}\int \d^4 x\sqrt{|g|}\, [-R+(\D\Phi)^2+2\K^2 V(\Phi)],
\]
where $\K^2 = 8\pi G$ and we take the scalar field $\Phi$ to be dimensionless.
For background solutions in which the evolution of the scalar field is (piece-wise)
{\it monotonic}, $\vphi(z)$ can be inverted to give $z(\vphi)$, allowing the Hubble rate $H=\dot{a}/a$ to be expressed in terms of some `fake superpotential' $W(\vphi)$ as
$
H(z) = - (1/2) W(\vphi).
$
The complete equations for the background are then
\[
\label{firstorder2}
\frac{\dot{a}}{a} = -\frac{1}{2}W, \quad 
\dot{\vphi} = W', \quad 2\eta \K^2 V = W'^2-\frac{3}{2} W^2,
\]
where $W' = \d W /\d \vphi$.
This first-order formalism goes back to the work of \cite{Salopek:1990jq} (for cosmology), 
where it was obtained by application of the Hamilton-Jacobi method. 
In \cite{Skenderis:2006jq} this formalism was linked to the notion of (fake) (pseudo-) supersymmetry.

The equations of motion for the perturbations are
\begin{align}
\label{pert_eom}
0&= \ddot{\zeta}+(3H+\dot{\ep}/\ep)\dot{\zeta}- \eta a^{-2} q^2\zeta, \nn \\
0&= \ddot{\gamma}_{ij}+3H\dot{\g}_{ij}-\eta a^{-2}q^2\g_{ij},
\end{align}
where $\vec{q}$ is the comoving wavevector of the perturbations,
and the background quantity $\ep(z)$ is defined as
$\ep = -\dot{H}/H^2 =2(W'/W)^2$.  
$\g_{ij}$ is a transverse traceless metric perturbation and 
$\zeta = \psi+(H/\dot{\vphi})\delta\vphi$
is the standard gauge-invariant variable constructed from a metric 
perturbation, $h_{ij} = -2\psi(z,\vec{x})\delta_{ij}$, and the scalar
perturbation $\delta\vphi$.

Defining now the analytically continued variables $\tK$ and $\tq$ according to
\[
\label{a/c}
\tK^2  = -\K^2, \quad \tq=-iq\, ,
\]
it is easy to see that a perturbed cosmological solution written in terms of the variables $\K$ and $q$ continues to a perturbed Euclidean domain-wall solution expressed in terms of the variables $\tK$ and $\tq$.

We have thus established that the correspondence between cosmologies
and domain-walls holds, not only for the background solutions, but
also for linear perturbations around them.  This is the basis for the
relation between power spectra and holographic 2-point functions, to
be discussed momentarily.  The argument can be generalized to
arbitrary order to relate non-Gaussianities to holographic
higher-point functions \cite{to_appear}.

\paragraph{Quantization of perturbations.} 

In the inflationary paradigm, cosmological perturbations 
originate on sub-horizon scales as quantum fluctuations of the vacuum.
Quantizing the perturbations in the usual manner, one finds the scalar and tensor 
superhorizon power spectra
\begin{align}
\Delta^2_S(q)& = \frac{q^3}{2\pi^2}\< \z(q)\z(-q)\>= \frac{q^3}{2\pi^2}|\z_{q(0)}|^2, \nn \\
\Delta^2_T(q) &= \frac{q^3}{2\pi^2}\< \g_{ij}(q)\g_{ij}(-q)\>=\frac{2 q^3}{\pi^2}|\g_{q(0)}|^2,
\end{align}
where $\g_{q(0)}$ and $\z_{q(0)}$ are the constant late-time values of the cosmological mode functions $\g_q(z)$ and $\z_q(z)$.

The mode functions are themselves solutions of the classical equations of motion (\ref{pert_eom}) 
(setting $\g_{ij}=\g_q e_{ij}$, for some time-independent polarization tensor $e_{ij}$).
To select a unique solution for each mode function we 
impose the Bunch-Davies vacuum condition $\zeta_q, \g_q \sim e^{-iq\t}$ 
as $\t \tt - \inf$, where the conformal time $\t = \int^z \d z'/a(z')$.
The normalization of each solution (up to an overall phase) may then be fixed by imposing the canonical commutation relations for the corresponding quantum fields.
This leads to the Wronskian conditions,
\[
 i = \z_q \Pi^{(\z)*}_q - \Pi^{(\z)}_q \z_q^*, \quad
i/2 = \g_q \Pi^{(\g)*}_q - \Pi^{(\g)}_q \g_q^*,
\]
where $\Pi^{(\zeta)}_q= 2 \ep a^3\K^{-2} \dot{\z_q}$ and $\Pi^{(\g)}_q = (1/4)a^3\K^{-2}\dot{\g}_q$ are the canonical momenta associated with each mode function, and we have set $\hbar$ to unity. 

To make connection with the holographic analysis to follow,
we introduce the linear response functions $E$ and $\Omega$ satisfying 
\[
\label{response_fns}
\Pi^{(\zeta)}_q=\Omega \,\zeta_q, \qquad  \Pi^{(\g)}_q= E \,\g_q.
\]
(These quantities are well-defined since we have already selected a unique solution for each mode function).
Substituting these definitions into the Wronskian conditions, 
which are valid at all times, the cosmological power
spectra may be re-expressed as
\[
\Delta^2_S(q) = \frac{-q^3}{4\pi^2 \Im \Omega_{(0)}(q)}, \quad
\Delta^2_T(q) = \frac{-q^3}{2\pi^2 \Im E_{(0)}(q)},
\]
where $\Im\Omega_{(0)}$ and $\Im E_{(0)}$ are the constant late-time values of the imaginary part of the response functions.
We will see shortly how the response functions also give the 2-point functions of the pseudo-QFT.

Let us now consider the corresponding domain-wall solution obtained by the applying the continuation (\ref{a/c}). 
The early-time behavior $\sim e^{-iq\t}$ of the cosmological perturbations maps to the exponentially decaying behavior $\sim e^{\tq \t}$ in the interior of the domain-wall ($\t \tt -\inf$).
Such regularity in the interior is a prerequisite for holography, explaining our choice of sign in the continuation of $q$.

The domain-wall response functions $\bar{E}$ and $\bar{\Omega}$ 
\cite{Papadimitriou:2004rz} are defined analogously to (\ref{response_fns}),
namely
\[
\bar{\Pi}^{(\z)}_{\tq} = -\bar{\Omega}\, \z_{\tq} \qquad
\bar{\Pi}^{(\g)}_{\tq} = -\bar{E}\, \g_{\tq},
\]
where $\bar{\Pi}^{(\z)}_{\tq} = 2\ep a^3\tK^{-2} \dot{\zeta}_{\tq}$ and 
$\bar{\Pi}^{(\g)}_{\tq} = (1/4)a^3\tK^{-2}\dot{\g}_{\tq}$
are the radial canonical momenta.  The minus signs are inserted so that
$\bar{\Omega}(-iq)=\Omega(q)$ and $\bar{E}(-iq)=E(q)$.
By choosing the arbitrary overall phase of the cosmological perturbations appropriately, we may ensure that the domain-wall perturbations are everywhere real.  The domain-wall response functions are then purely real, while their cosmological counterparts are complex. 

\paragraph{Holographic analysis.}

There are two classes of domain-wall solutions for which holography
is well understood.

\paragraph{(i) Asymptotically AdS domain-walls.}

In this case the solution behaves asymptotically as
\[
a(z) \sim e^{z}, \qquad \vphi \sim 0 \qquad {\rm as} \quad z \to \infty.
\]
The boundary theory has a UV fixed point which corresponds to the bulk AdS critical point.
Depending on the rate at which $\vphi$ approaches zero as $ z \to \infty$,
the QFT is either a deformation of the CFT or else the CFT in a state in which the dual scalar operator acquires a nonvanishing vacuum expectation value (see \cite{Skenderis:2002wp} for details). 
Under the domain-wall/cosmology correspondence, these solutions are mapped to
cosmologies that are asymptotically de Sitter at late times.

\paragraph{(ii) Asymptotically power-law solutions.}

In this case the solution behaves asymptotically as
\[
\label{power_law}
a(z) \sim (z/z_0)^n, \quad \vphi \sim \sqrt{2 n} \log (z/z_0) \quad {\rm as} \quad z \to \infty,
\]
where $z_0=n{-}1$.
This case has only very recently been understood \cite{Kanitscheider:2008kd}. 
 For $n=7$ the asymptotic geometry is the near-horizon limit of a stack of D2 brane solutions. 
In general, these solutions describe QFTs with a dimensionful coupling
constant in the regime where the dimensionality of the coupling constant drives the dynamics.
Under the domain-wall/cosmology correspondence, these solutions are mapped to cosmologies that 
are asymptotically power-law at late times. 

Holographic 2-point functions are now obtained by solving the linearized equations of motion about the domain-wall solution with Dirichlet boundary conditions at infinity and imposing regularity in the interior.  It will suffice to discuss the 2-point function for the energy-momentum tensor.
On general grounds, the 2-point function takes the form
\[
\label{ABdef}
\<T_{ij}(\tq)T_{kl}(-\tq)\> = A(\tq)\Pi_{ijkl}+B(\tq)\pi_{ij}\pi_{kl},
\]
where $\Pi_{ijkl}$ is the three-dimensional transverse traceless projection operator defined by
$\Pi_{ijkl}= \pi_{i(k}\pi_{l)j}{-}(1/2)\pi_{ij}\pi_{kl}$ and 
$\pi_{ij} = \delta_{ij} {-} \bar{q}_i \bar{q}_j /\bar{q}^2$.
The holographic computation amounts to the extracting the coefficients $A$ and $B$ from the asymptotics of the linearized solution. Using the radial Hamiltonian method developed in \cite{Papadimitriou:2004rz}, one finds
\[
\label{ABeqns}
A(\tq) =  4 \bar{E}_{(0)}(\tq) , \quad
B(\tq) = \frac{1}{4} \bar{\Omega}_{(0)}(\tq),
\]
where the zero subscript indicates the leading (constant) term as $z \tt \inf$, after holographic renormalization has been performed \cite{Papadimitriou:2004rz, proceedings}.

\paragraph{Continuation to the pseudo-QFT.}

We now wish to re-express the bulk analytic continuation (\ref{a/c}) in terms of QFT variables.
This amounts to
\[
\bar{N}^2 = - N^2,  \qquad \tq = -i q,
\]
where the barred quantities are associated with the QFT dual to the domain-wall.
We thus find that the power spectrum for any inflationary cosmology that is 
asymptotically de Sitter or asymptotically power law can be 
directly computed from the 2-point function of a three-dimensional QFT
via the formulae:
\[
\label{result}
\Delta^2_S(q) = \frac{-q^3}{16 \pi^2 \Im B(-i q)}, \ \ \ 
\Delta^2_T(q) = \frac{-2 q^3}{\pi^2 \Im A(-i q)}.
\]

\paragraph{Beyond the weak gravitational description.}

In the discussion so far we have assumed that the description in terms
of gravity coupled to a scalar field is valid at early times, and that
the perturbative quantization of fluctuations can be justified.  The
holographic description also allows us to obtain results when these
assumptions do not hold.  At early times, the theory may be strongly
coupled with no useful description in terms of low-energy fields (such
as the metric and the scalar field).  The holographic set-up allows us
to extract the late-time behavior of the system, which can be
expressed in terms of low-energy fields, from QFT correlators. 
(For further details see \cite{proceedings}).
Here we only note that this is the counterpart of the standard 
fact that in gauge/gravity duality
the asymptotic behavior of bulk fields near the boundary of
spacetime is reconstructed by the correlators of the dual QFT 
\cite{de Haro:2000xn}. This late-time behavior is
precisely the information we need to compute the primordial power
spectra and other cosmological observables.

Ideally one would deduce from a string/M-theoretic construction what
the dual QFT is. Instead we initiate here a holographic
phenomenological approach. The dual QFT would involve scalars,
fermions and gauge fields and it should admit a large $N$ limit. The
question is then whether one can find a theory which is compatible
with current observations.  In particular, one might consider either
deformations of CFTs or theories with a single dimensionful parameter,
as these QFTs have already featured in our discussion above.

\begin{figure}[t]
\center
\hspace{-2pc}
\includegraphics[width=6cm]{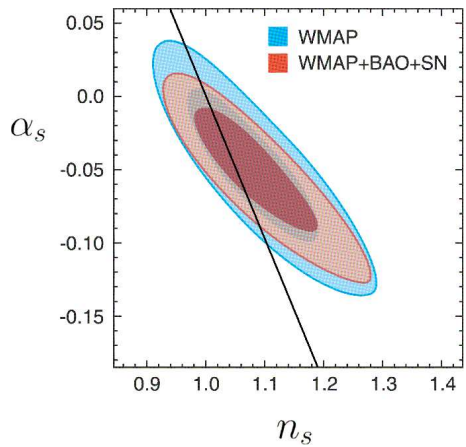} 
\caption{The straight line is the leading order prediction of 
holographic models with a single dimensionful coupling constant for the 
correlation of the running $\alpha_s$ and the scalar tilt $n_s$. 
The data show the $68\%$ and $95\%$ CL constraints (marginalizing over tensors)
at $q = 0.002\, \mathrm{Mpc}^{-1}$, and are taken from Fig.~4 of \cite{Komatsu:2008hk}.
As new data appear the allowed region should shrink to a point,
which is predicted to lie close to the line.}
\end{figure}

We will discuss here super-renormalizable theories that contain one
dimensionful coupling constant.  A prototype example is $SU(N)$
Yang-Mills theory coupled to a number of scalars and fermions, all in
the adjoint of $SU(N)$.  To extract predictions we need to compute the
coefficients $A$ and $B$ of the 2-point function of the
energy-momentum tensor (\ref{ABdef}), analytically continue the
results and insert them in the holographic formulae for the power
spectra.

Firstly, the leading contribution to the 2-point function of the energy-momentum tensor is at
one loop. Since the energy-momentum tensor has dimension three,
\[
A(\bar{q}) \sim \bar{N}^2 \bar{q}^3, \qquad  B(\bar{q}) \sim \bar{N}^2 \bar{q}^3.
\]
A generic such model thus leads to a scale-invariant spectrum at leading order in $N$.
To fix the parameters of these models we may then compare with observations.
Comparing the observed amplitude of the scalar power spectrum \cite{Komatsu:2008hk} with its holographic value we find $N \sim O(10^4)$, justifying the large $N$ limit. 
To determine the coupling constant $g_{\mathrm{YM}}^2$ we may compare with the tilt of the spectrum.  The precise formula requires a two-loop computation  \cite{footnote2} and will be reported elsewhere \cite{to_appear}. One can obtain an order
of magnitude estimate, however, on general grounds. The perturbative expansion depends on the effective dimensionless coupling constant $g_{\mathrm{eff}}^2 = g_{\mathrm{YM}}^2 \bar{N}/\tq$, and the leading correction to the 2-point
function yields
$
(n_s{-}1) = c g_{\mathrm{eff}}^2,
$ 
where the constant $c$ is of order one and depends on the details of
the theory.  From Table 4 of \cite{Komatsu:2008hk} one finds that
$(n_s{-}1) \sim O(10^{-2})$, thus we find that $g_{\mathrm{eff}}^2 \sim
O(10^{-2})$ justifying the perturbative QFT treatment.
Independently of the details of the theory,
the scalar index runs: $\alpha_s = \d n_s/\d\ln q =  -(n_s{-}1) 
+ O(g_{\mathrm{eff}}^4)$. This prediction is qualitatively different from slow-roll inflation
(for which $\alpha_s/(n_s{-}1)$ is of first-order in slow-roll \cite{Kosowsky:1995aa}), yet is nonetheless consistent with the constraints on $n_s$ and $\alpha_s$ given in \cite{Komatsu:2008hk} for a wide range of values of $n_s$ and
$\alpha_s$, as illustrated in Fig.~2. 
The ratio of tensor to
scalar power spectra can be computed from (\ref{result}),
yielding $r = 32 \Im B(-iq)/\Im A(-iq)$.
For massless scalars and for vector fields $A = B = (1/256)\bar{N}^2 \tq^3$ 
(for conformally coupled scalars $B=0$ instead), 
and for massless fermions $A = (1/128) \bar{N}^2 \tq^3$ and $B=0$. 
With appropriate field content one can thus satisfy the current 
observational bound on $r$.

Once $N$, $g_{\mathrm{YM}}^2$ and the field content are fixed, all
other cosmological observables (such as non-Gaussianities, {\it
 etc.})~follow uniquely from straightforward computations.
We will present details of
the correspondence between higher-order QFT correlation functions and
non-Gaussian cosmological observables elsewhere \cite{to_appear}.  Our
results indicate, however, that the non-Gaussianity parameter
$f_{NL}^{\mathrm{local}}$ \cite{Komatsu:2001rj} is independent of $N$
to leading order, consistent with current observational evidence
\cite{Komatsu:2008hk}.

\paragraph{Conclusions.} 

We have presented a concrete proposal describing holography for cosmology,
and initiated a holographic phenomenological approach 
capable of satisfying current observational constraints.
Clearly one would like to further develop holographic phenomenology
and obtain precise predictions for the cosmological observables to be 
measured by forthcoming experiments.
We hope to report on this in the near future.

\nopagebreak

{\it Acknowledgements.}  We thank NWO for support.


\begin{thebibliography}{99}

\bibitem{Komatsu:2008hk}
E.~Komatsu et al.~(WMAP), Astrophys.~J.~Suppl.~{\bf 180}, 330 (2009), arxiv:0803.0547.
%%CITATION = 0803.0547;%%

\bibitem{Maldacena:2002vr}
J.~Maldacena, JHEP {\bf 05}, 013 (2003), astro-ph/0210603.
%%CITATION = ASTRO-PH/0210603;%%

\bibitem{Cvetic:1994ya}
M.~Cvetic and H.~H.~Soleng, Phys.~Rev.~{\bf D51}, 5768 (1995), hep-th/9411170.
%%CITATION = HEP-TH/9411170;%%

\bibitem{Skenderis:2006jq}
K.~Skenderis and P.~K.~Townsend, Phys.~Rev.~Lett.~{\bf 96}, 191301 (2006), hep-th/0602260;
J.~Phys.~{\bf A40}, 6733 (2007), hep-th/0610253.
%%CITATION = HEP-TH/0602260;%%
%%CITATION = HEP-TH/0610253;%%

\bibitem{footnote1}
This era should then be followed
by a hot big bang cosmology, as in standard discussions. Here we 
only discuss the very early universe, \iec the times when the primordial
cosmological perturbations were generated (the inflationary epoch).

\bibitem{Bergshoeff:2007cg}
E.~A.~Bergshoeff, J.~Hartong, A.~Ploegh, J.~Rosseel and D.~van den Bleeken,
JHEP {\bf 07}, 067 (2007), arxiv:0704.3559;
K.~Skenderis, P.~K.~Townsend and A.~van Proeyen, JHEP {\bf 08}, 036 (2007), arxiv:0704.3918.
%%CITATION = 0704.3559;%%
%%CITATION = 0704.3918;%%

\bibitem{Skenderis:2008dh}
K.~Skenderis and B.~C.~van Rees, Phys.~Rev.~Lett.~{\bf 101}, 081601 (2008), arxiv:0805.0150.
%%CITATION = 0805.0150;%%

\bibitem{Salopek:1990jq}
D.~S.~Salopek and J.~R.~Bond, Phys.~Rev.~{\bf D42}, 3936 (1990).
%%CITATION = PHRVA,D42,3936;%%

\bibitem{to_appear}
P.~McFadden and K.~Skenderis, to appear.

\bibitem{Papadimitriou:2004rz}
I.~Papadimitriou and K.~Skenderis, JHEP {\bf 10}, 075 (2004), hep-th/0407071.
%%CITATION = HEP-TH/0407071;%%

\bibitem{Skenderis:2002wp}
K.Skenderis, Class.~Quant.~Grav.~{\bf 19}, 5849 (2002), hep-th/0209067.
%%CITATION = HEP-TH/0209067;%%

\bibitem{Kanitscheider:2008kd}
I.~Kanitscheider, K.~Skenderis and M.~Taylor, JHEP {\bf 09}, 094 (2008), arxiv:0807.3324.
%%CITATION = 0807.3324;%%

\bibitem{proceedings}
P.~McFadden and K.~Skenderis, arxiv:1001.2007.
%%CITATION = 1001:2007;%%

\bibitem{de Haro:2000xn}
  S.~de Haro, S.~N.~Solodukhin and K.~Skenderis,
  Commun.~Math.~Phys.~{\bf 217}, 595 (2001), hep-th/0002230.
%%CITATION = HEP-TH/0002230;%%

\bibitem{Komatsu:2001rj}
E.~Komatsu and D.~N.~Spergel, Phys.~Rev.~{\bf D63}, 063002 (2001), astro-ph/0005036.
%%CITATION = ASTRO-PH/0005036;%%

\bibitem{footnote2}
Super-renormalizable theories have infrared divergences, but large $N$ resummation leads to well-defined expressions with $g_{\mathrm{YM}}^2$ effectively playing the role of an infrared regulator. The exact amplitudes are nonanalytic functions of the coupling constant \cite{Jackiw:1980kv}.
Note that our analytic continuation to pseudo-QFT does not involve the coupling constant.

\bibitem{Jackiw:1980kv}
R.~Jackiw and S.~Templeton, Phys.~Rev.~{\bf D23}, 2291 (1981); 
T.~Appelquist and R.~D.~Pisarski, Phys.~Rev.~{\bf D23}, 2305 (1981).
%%CITATION = PHRVA,D23,2291;%%
%%CITATION = PHRVA,D23,2305;%%

\bibitem{Kosowsky:1995aa}
A.~Kosowsky and M.~Turner, Phys.~Rev.~{\bf D52}, 1739 (1995), astro-ph/9504071.
%%CITATION = ASTRO-PH/9504071;%%

\end{thebibliography}
\end{document}